\documentclass[a4paper,11pt]{article}
\usepackage{pos}
\usepackage{siunitx}
\usepackage{xspace}
\usepackage{wrapfig}

\newcommand{\reffig}[1]{Fig.~\ref{#1}}

\newcommand{\refsec}[1]{Section~\ref{#1}}

\newcommand{\refref}[1]{Ref.~\cite{#1}}

\newcommand{\sibyll}{Sibyll~2.3c\xspace}
\newcommand{\epos}{EPOS-LHC\xspace}
\newcommand{\qgsjet}{QGSJet-II.04\xspace}
\newcommand{\dpmjet}{DPMJet-III 19.1\xspace}
\newcommand{\corsika}{\textsc{Corsika}\xspace}
\newcommand{\fluka}{\textsc{Fluka}\xspace}
\newcommand{\mceq}{\textsc{MCEq}\xspace}

\title{Sensitivity of seasonal variations of the muon energy spectrum in air showers to hadronic interaction models and the cosmic-ray mass composition}
\ShortTitle{Seasonal variations of the muon energy spectrum in air showers}

\author*[a]{Stef Verpoest}
\author[a]{Marisol Catalan Olais}
\author[a]{Frank Schroeder}

\affiliation[a]{Bartol Research Institute, Department of Physics and Astronomy, University of Delaware,\\
 Sharp Lab, 104 The Green, Newark DE, 19716, United States of America}

\emailAdd{verpoest@udel.edu}

\abstract{The energy spectrum of muons produced in air showers depends not only on the properties of the primary particle, but also on the atmosphere. This is because of the competition between decay and interaction of the parent mesons, which depends on the atmospheric density. As a result, the number of muons at ground shows a seasonal variation, with the strength of the variation depending on the primary cosmic-ray energy, mass, as well as the energy of the detected muons. In this contribution, we study the variations of the muon energy spectrum in air showers using MCEq, a numerical solver of the cascade equations. In particular, we show how the amplitude of the seasonal variations of the number of high-energy ($\mathcal{O}$(TeV)) muons in the shower depends on the primary energy, mass, as well as the assumed hadronic interaction model. A measurement of these variations is possible with the combination of a surface air-shower array and a deep underground detector, which provide a simultaneous measurement of the primary energy and the high-energy muon multiplicity, and may provide a new way to probe the cosmic-ray mass composition and hadronic interactions.}

\ConferenceLogo{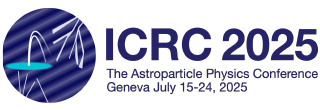}

\FullConference{39th International Cosmic Ray Conference (ICRC2025)\\
 15–24 July 2025\\
Geneva, Switzerland\\}
\begin{document}
\maketitle

\section{Introduction}

Cosmic rays interacting in the Earth's atmosphere produce large particle cascades known as extensive air showers. The air showers contain muons which are dominantly produced in the decays of mesons produced in the hadronic cascade. Muons arriving at the Earth's surface as part of an air shower span a large range of energies and production altitudes. For air shower experiments consisting of a surface array of particle detectors, mostly the large numbers of muons with $E_\mu \approx \mathcal{O}(\SI{100}{\mega\eV}-\SI{1}{\giga\eV})$ are relevant, while deep underground detectors only observe muons with $E_\mu \gtrsim \mathcal{O}(\SI{100}{\giga\eV}-\SI{1}{\tera\eV})$.

The production of muons depends on the local density of the atmosphere and therefore shows a seasonal effect as the atmosphere warms and cools throughout the year. The seasonal variation of the muon rate in underground detectors, which relates to the inclusive atmospheric muon flux, has been described and observed in detail (see for example \refref{Verpoest:2024dmc} for a recent overview of the phenomenology and \refref{OPERA:2018jif} for a compilation of measurements from different experiments). In contrast, also the variations of muons in air showers at specific primary energies are of interest, but have not been described in detail.

In this contribution, we explore the atmospheric dependence of muon production in air showers using the Matrix Cascade Equations (\mceq) code~\cite{Fedynitch:2015zma}, a numerical solver of cascade equations of particle numbers in the atmosphere. As will be shown, the size of the variations increases with muon energy over a large range of energies. Therefore, we will discuss in particular the variations of the TeV muon multiplicity in air showers, and its dependence on hadronic interaction models and the cosmic-ray mass composition. This is of interest for surface air shower arrays combined with deep underground detectors: by observing an air shower at the surface and determining its primary energy, combined with a measurement of high-energy muons from the same shower in the deep detector, the seasonal variations of the high-energy muon content can be determined as a function of the primary energy. This is possible, for example, at the IceCube Neutrino Observatory~\cite{IceCube:2016zyt}, which combines a deep in-ice detector with a surface array, and has already measured the yearly-average multiplicity of TeV muons in air showers in the \SI{2.5}{\peta\eV} to \SI{100}{\peta\eV} range~\cite{IceCube:2025baz}.

\section{Spectrum of muons in air showers}

\begin{figure}
    \centering
    \includegraphics[width=0.49\linewidth]{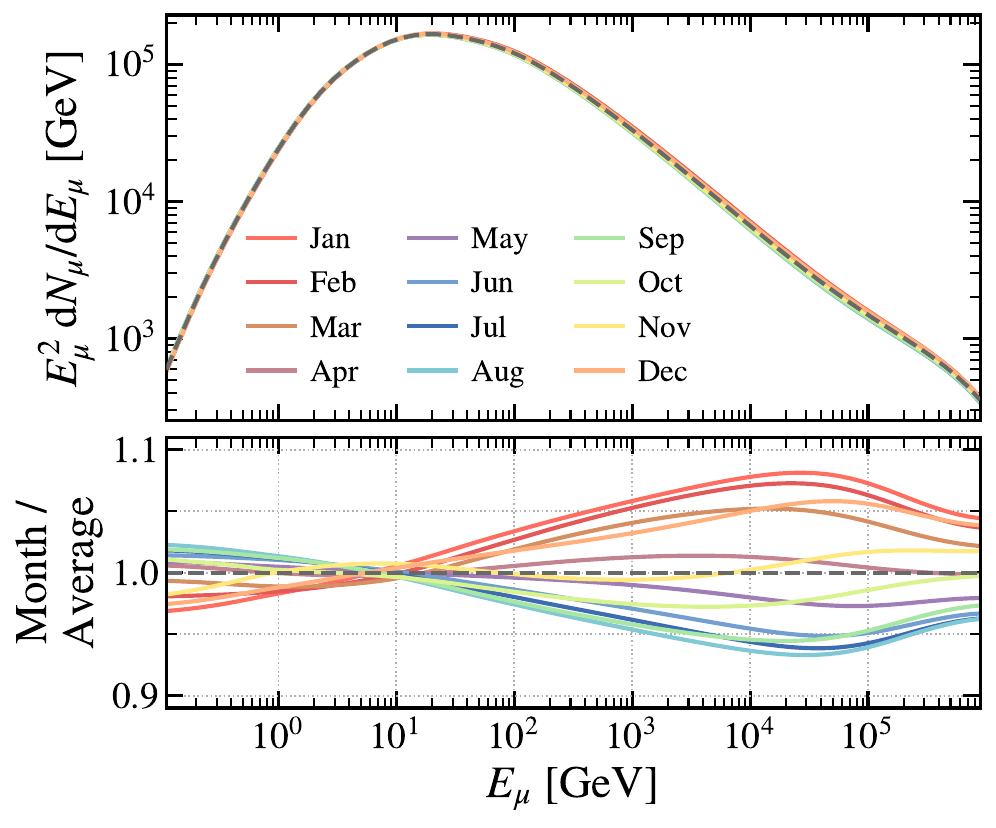}
    \caption{Energy spectrum of muons in vertical air showers initiated by \SI{10}{\peta\eV} protons, calculated with \mceq using different South Pole atmospheres.}
    \label{fig:monthly_spectra}
\end{figure}

A comparison of the muon energy spectrum in air showers expected for different atmospheric profiles is shown in \reffig{fig:monthly_spectra}. The spectra were calculated using \mceq, solving the atmospheric cascade equations for a single primary particle, and they therefore represent the expected average spectra over many air showers. We use a vertically down-going \SI{10}{\peta\eV} proton as the primary particle, an observation altitude of \SI{2835}{\m}, and South Pole atmospheric density profiles from \refref{DeRidder:2019ofg}, as relevant for the case of IceCube. These parameters will be used throughout the work, assuming \sibyll~\cite{Riehn:2017mfm} for the high-energy hadronic interaction model, unless indicated otherwise.

A clear difference in the shape of the muon spectra from month to month can be seen. The effect is small for the low-energy muons dominating at ground, showing a slightly decreased number of muons in the warmer months and vice versa for the colder months. A   bove \SI{10}{\giga\eV}, the opposite is true; warmer months have more such muons, and the effect gets stronger with increasing muon energy. This can be understood from the growing decay lengths of parent mesons, mainly charged pions and kaons, which become similar to their interaction lengths at these energies~\cite{Gaisser_Engel_Resconi_2016, Fedynitch:2018cbl}. As a result, changes in the atmospheric density are reflected more strongly in the muon number, as decay is no longer the dominant scenario for the mesons. At higher energies, around \SI{30}{\tera\eV}, the seasonal effect becomes smaller again as a result of the increasing fraction of muons from prompt decays of charmed mesons and unflavored vector mesons.

\begin{figure}
    \centering
    \includegraphics[width=0.47\linewidth]{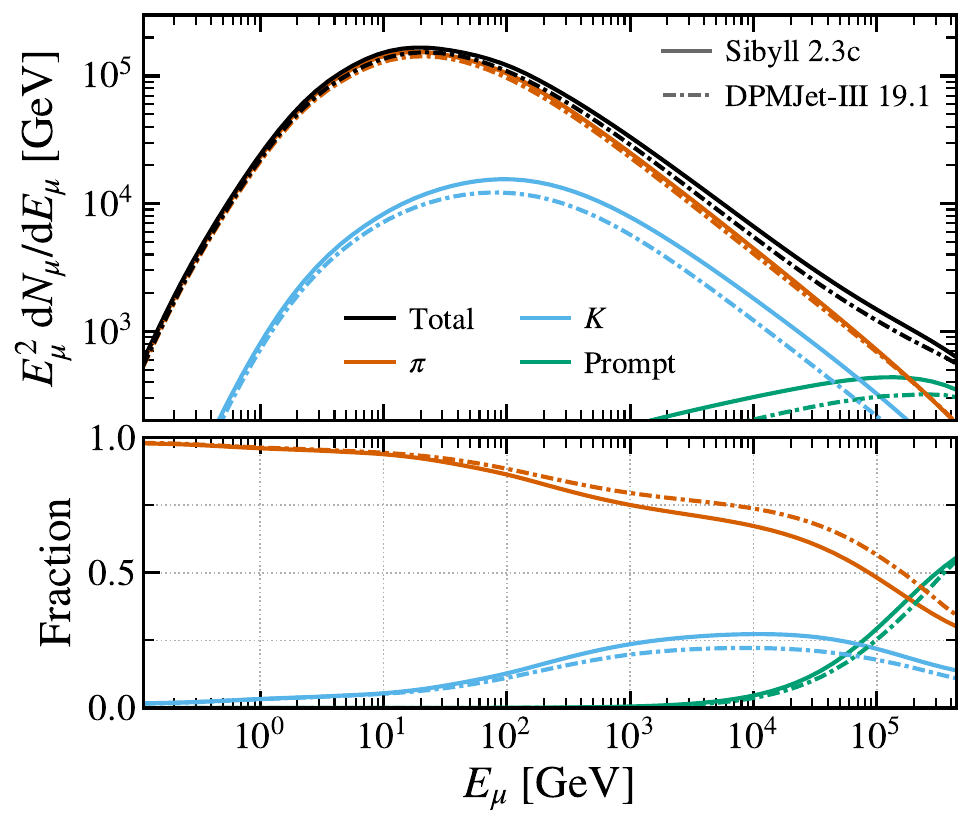}\hfill\includegraphics[width=0.47\linewidth]{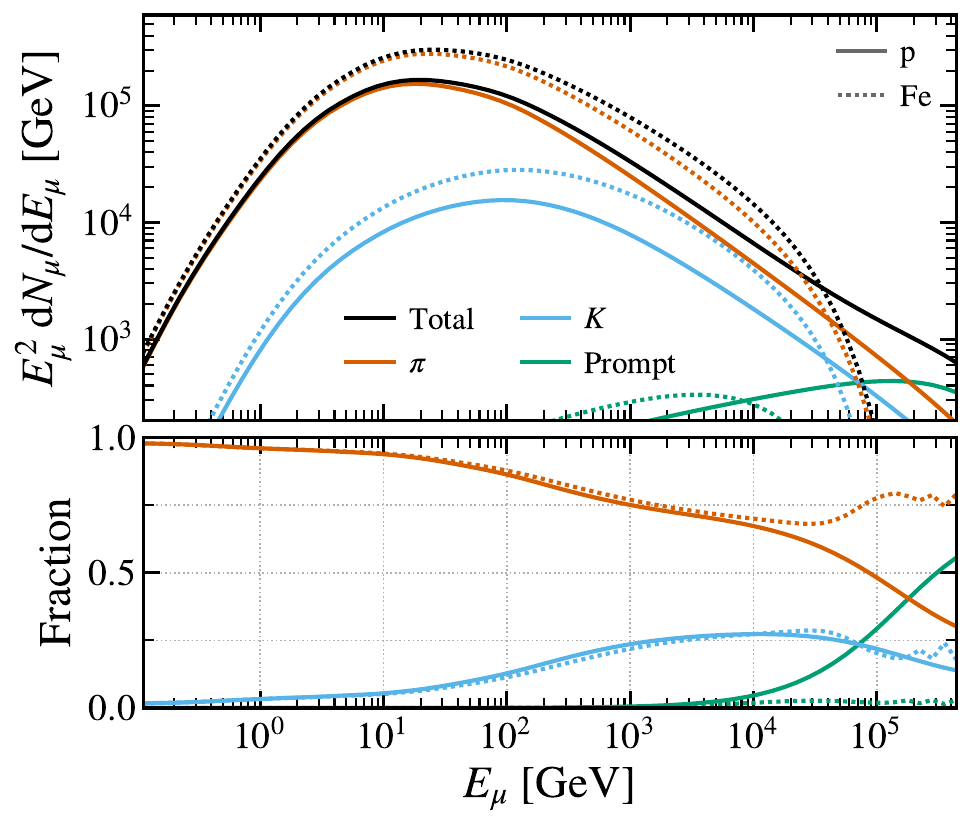}
    \caption{Contributions from different production channels to the muon energy spectrum in \SI{10}{\peta\eV} proton showers. Left: Comparison between the hadronic models \sibyll and \dpmjet. Right: Comparison between air showers initiated by protons and iron nuclei. Iron cuts off at high energies because the shower develops with an energy per nucleon of 10 PeV / 56.}
    \label{fig:spectrum_contributions}
\end{figure}

The contributions from different production channels to the total muon spectrum in the air shower are shown in \reffig{fig:spectrum_contributions}. The left panel shows a comparison between predictions based on the hadronic interaction models \sibyll and \dpmjet~\cite{Fedynitch:2015kcn}. These models were chosen for the comparison plot as they show the largest difference in the variations of the high-energy muon number of all the models considered in this work, as will be shown in \refsec{sec:seasonalTeV}. As expected, muons dominantly originate from pion decay for all models, except above several \SI{100}{\tera\eV}, where the prompt component takes over. The contribution from kaons grows from close to zero at the lowest energies to about 25\% at $E_\mu = \SI{1}{\tera\eV}$. The right panel of \reffig{fig:spectrum_contributions} compares the spectra in proton and iron showers. As expected, iron showers have more muons over most of the energy range, except for the observed cutoff at higher energies resulting from the fact that for iron showers the primary energy is shared between 56 nucleons (superposition model).

\begin{figure}
    \centering
    \includegraphics[width=0.49\linewidth]{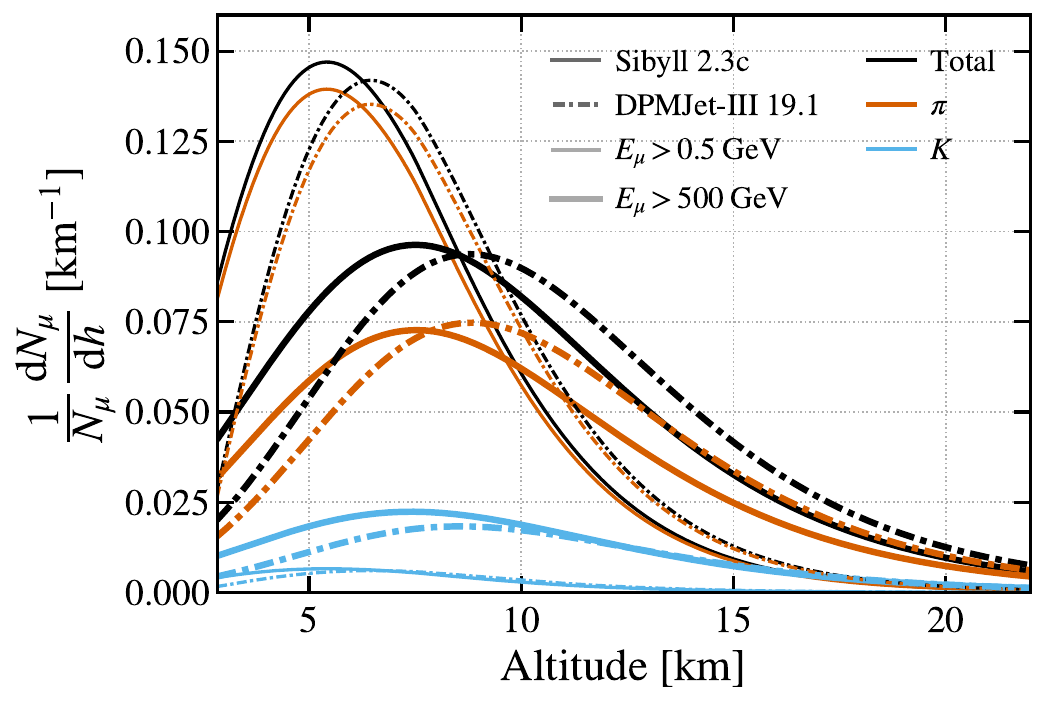}\hfill\includegraphics[width=0.49\linewidth]{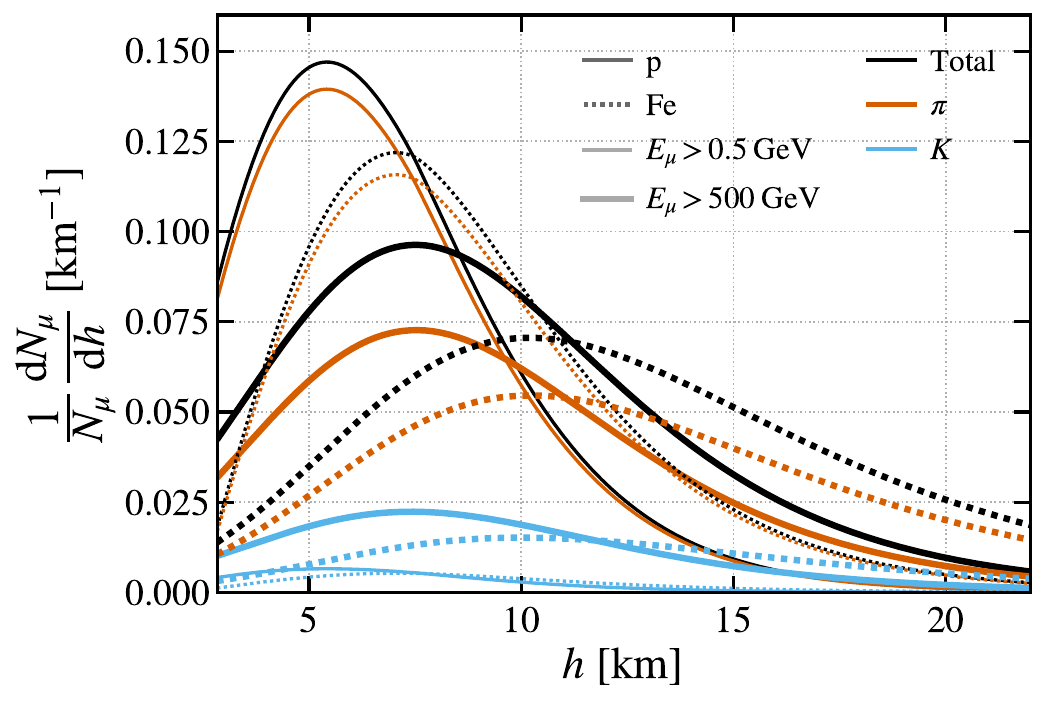}
    \caption{Longitudinal production profiles as a function of altitude for all muons above \SI{0.5}{\giga\eV} (narrow line) and above \SI{500}{\giga\eV} (wide line) in vertical \SI{10}{\peta\eV} proton showers. The profiles are plotted down to an altitude of \SI{2835}{\m} a.s.l., corresponding to the surface level of the ice sheet at the South Pole. Left: Comparison between the hadronic models \sibyll and \dpmjet. Right: Comparison between air showers initiated by protons and iron nuclei.}
    \label{fig:production_pdf}
\end{figure}

In \reffig{fig:production_pdf}, the derivative of the longitudinal muon profiles with respect to altitude are shown. Ignoring muon decay, these represent the longitudinal production profiles as a function of the altitude. The production of muons with $E_\mu > \SI{0.5}{\giga\eV}$ peaks at lower altitudes than the muons with $E_\mu > \SI{500}{\giga\eV}$, with the exact altitude depending on the hadronic interaction model, as shown in the left panel. The right panel shows that, as expected, iron showers develop on average earlier in the atmosphere than proton showers. The effect of these differences on possible observations will be discussed in \refsec{sec:seasonalTeV}.

\begin{figure}
    \centering
    \includegraphics[width=0.465\linewidth]{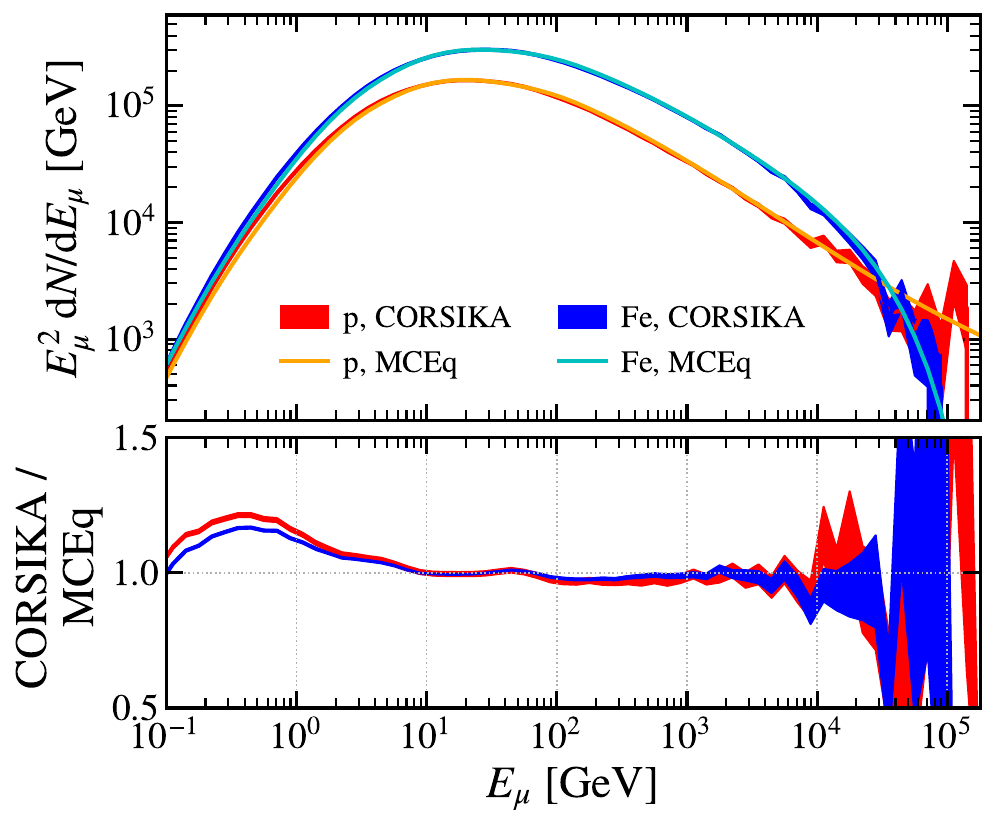}\hfill\includegraphics[trim=0 0.3cm 0 0, clip, width=0.46\linewidth]{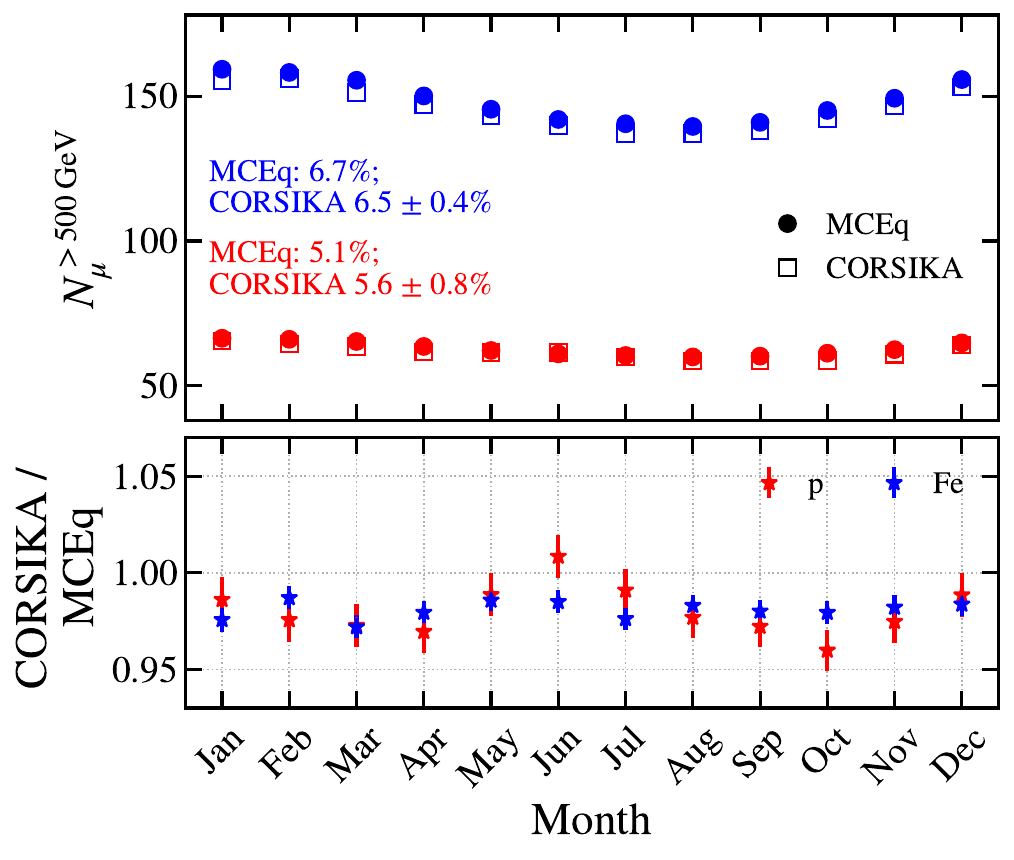}
    \caption{Comparison between \mceq calculations and the average of 1000 \corsika simulations for vertical \SI{10}{\peta\eV} showers. The \mceq calculations use \sibyll as the high-energy hadronic interaction model with a smooth transition to \dpmjet below \SI{100}{\giga\eV}; the \corsika simulations use Sibyll~2.3d for interactions above \SI{80}{\giga\eV} and \fluka below. Left: Muon energy spectrum for proton and iron showers. Right: Monthly variation in the high-energy muon number ($E_\mu > \SI{500}{\giga\eV}$). The text in the upper panel gives the amplitude of the relative variation.}
    \label{fig:corsika}
\end{figure}

In addition to the \mceq calculations, a set of Monte-Carlo simulations of air showers was produced using \corsika~\cite{Heck:1998vt} for comparison.
% The \sibyll \mceq calculations to the average of 1000 air showers simulated at the same primary energy, mass, and direction using Sibyll~2.3d for the high-energy hadronic interaction model. 
% The \corsika simulations use FLUKA as the low-energy hadronic model for interactions below \SI{80}{\giga\eV}, while \mceq transitions to \dpmjet below \SI{100}{\giga\eV}.
The resulting muon spectra are compared in the left panel of \reffig{fig:corsika}. There is some discrepancy between the spectra, especially below $E_\mu = \SI{10}{\giga\eV}$, which may result from the different low-energy hadronic models used in both calculations (details in caption) and should be investigated for future work. At higher muon energies, the difference is smaller and shows less dependence on the energy. The right panel shows a comparison of the total number of muons with $E_\mu > \SI{500}{\giga\eV}$. While there is an offset in the absolute muon number, the seasonal variation throughout the year has an amplitude which is consistent between \corsika and \mceq within the statistical uncertainty. For a future measurement of this observable, discussed in the next section, a more detailed comparison should be performed.

\section{Seasonal variation of the TeV muon number}\label{sec:seasonalTeV}

\begin{figure}
    \centering
    \includegraphics[width=0.47\linewidth]{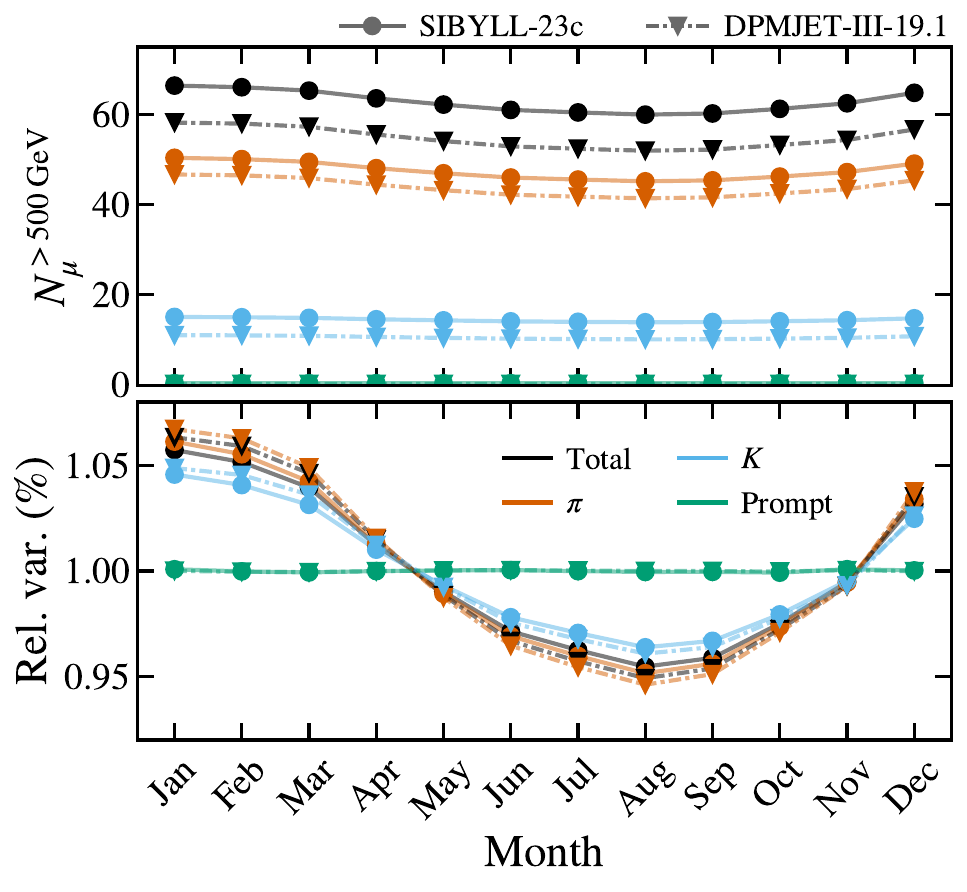}\hfill\includegraphics[width=0.47\linewidth]{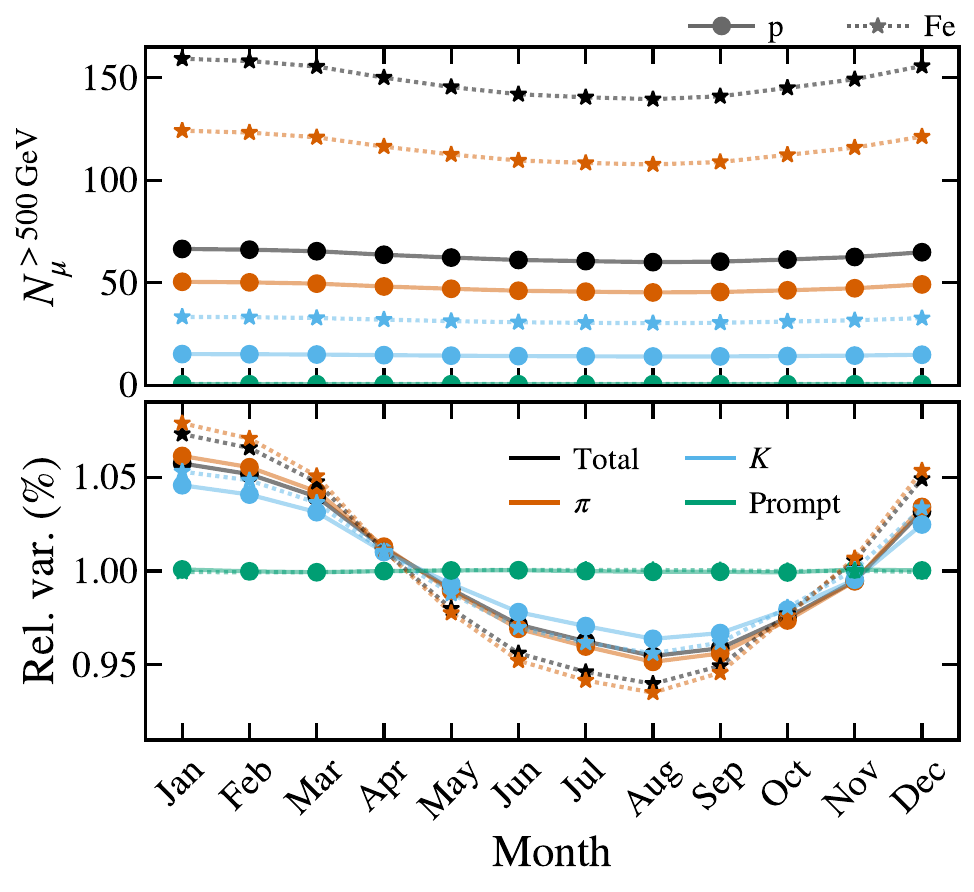}
    \caption{Variation in the number of muons with energy larger than \SI{500}{\giga\eV} throughout the year in vertical showers initiated by \SI{10}{\peta\eV} proton primaries. The top panel shows the total number of muons and contributions from different channels. The bottom panel shows the relative variation in each component compared to the yearly average. Left: Comparison between the hadronic models \sibyll and \dpmjet. Right: Comparison between air showers initiated by protons and iron nuclei.}
    \label{fig:relvar_contributions}
\end{figure}

In this section, we focus on the total number of muons with energies above \SI{500}{\giga\eV} in air showers , $N_\mu^{> \SI{500}{\giga\eV}}$ (``TeV muons''). As was shown in \reffig{fig:monthly_spectra}, a significant seasonal variation is expected in this observable. \reffig{fig:relvar_contributions} shows this variation between different months for the total number of muons in the shower, as well as for the contributions from different production channels separately. The bottom panel shows the relative variation, i.e. the ratio of the muon number in one month to the yearly average.
The number of muons produced by pions shows a larger relative variation than that from kaons, which can be understood as a result of the comparatively shorter decay length of charged kaons. The relative variation of the total number of muons has an amplitude in between.

\begin{figure}
    \centering
    \includegraphics[width=0.49\linewidth]{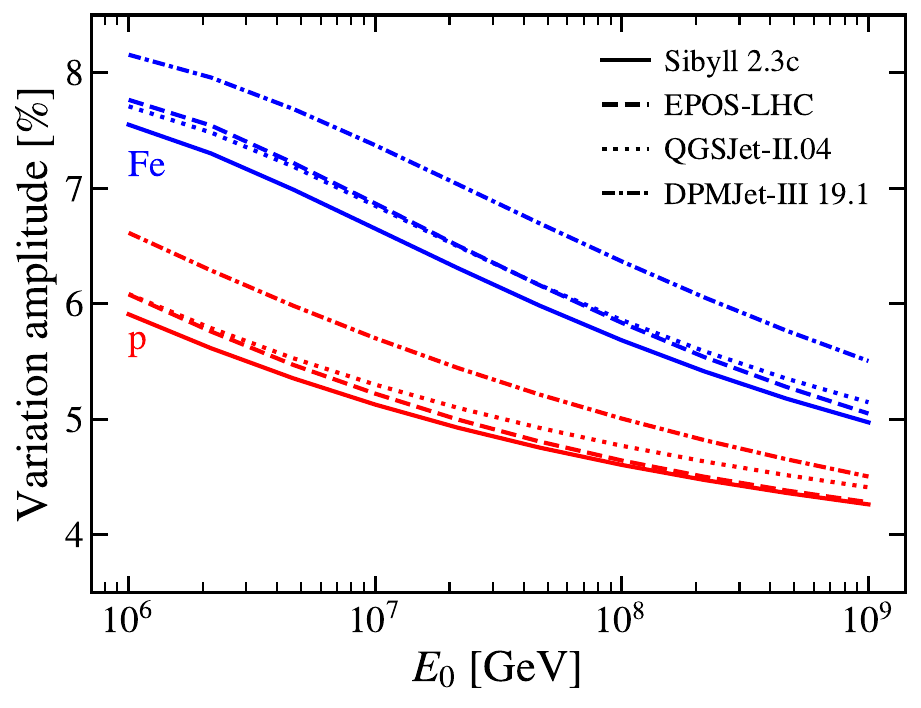}
    \caption{Amplitude of the relative seasonal variations expected in the number of muons with energy larger than \SI{500}{\giga\eV} in vertical air showers with primary energy $E_0$. Calculations were performed with \mceq using different hadronic interaction models.}
    \label{fig:relvar_all}
\end{figure}

We calculated the expected amplitude of the relative variation in the high-energy muon number as a function of primary cosmic-ray energy between \SI{1}{\peta\eV} and \SI{1}{\exa\eV} for vertical proton and iron primaries using \mceq and the hadronic interaction models \sibyll, \dpmjet, \qgsjet~\cite{Ostapchenko:2010vb}, and \epos~\cite{Pierog:2013ria}. The amplitude of the relative variation is defined as the mean of the extremal relative deviations from the yearly average, i.e.
% \begin{equation}
%     A_\mathrm{RV} = \frac{\max(RV_\mathrm{month}) - \min(RV_\mathrm{month})}{2}
% \end{equation}
\begin{equation}
    A_\mathrm{Rel.~Var.} = \frac{\max(N_{\mu, \mathrm{month}}^{> \SI{500}{\giga\eV}}) - \min(N_{\mu, \mathrm{month}}^{> \SI{500}{\giga\eV}})}{2 \langle N_\mu^{> \SI{500}{\giga\eV}}\rangle}.
\end{equation}
The results are shown in \reffig{fig:relvar_all}. \sibyll predicts the smallest seasonal variations, while \dpmjet predicts the largest. This effect can, on the one hand, be understood as a result of the larger contribution of pions compared to kaons in \dpmjet, as was shown in \reffig{fig:spectrum_contributions}. However, comparing the relative variations of muons from the different channels, as in \reffig{fig:relvar_contributions}, we see that both the pion and kaon components show a larger variation amplitude in \dpmjet compared to \sibyll. This can be understood from the higher production altitude for both components in \dpmjet, which was shown in \reffig{fig:production_pdf}, as the seasonal variation in the atmospheric density is larger in this part of the atmosphere compared to lower altitudes (see e.g. \refref{DeRidder:2019ofg}). The larger variations predicted for iron primaries compared to protons can similarly be understood through the earlier development of iron showers in the atmosphere, while there is only little difference in the relative contributions from pions and kaons in the relevant part of the energy spectrum. The decrease of the variation amplitude with increasing shower energy results from the fact that \SI{500}{\giga\eV} muons will be produced later in the development of higher-energy showers, i.e. at lower altitudes, where atmospheric variations are smaller.

A measurement of this effect could be performed with an air shower array combined with a deep underground detector, such as the IceCube Neutrino Observatory. Air showers in a specific primary energy range can be selected with the surface detector, while the average multiplicity of high-energy muons in these showers can be determined with the underground detector. As the relative variation is determined as a ratio of two measurements, various detector systematic uncertainties may cancel. This provides a new way for such detector setups to probe the cosmic-ray mass composition as a function of primary energy and to test predictions from hadronic interaction models. As the amplitude of the seasonal variation seems largely driven by the depth of the muon production, a measurement may even be able to constrain interaction parameters such as various hadronic cross sections~\cite{Hymon}.

\section{Conclusion \& Outlook}

The production of muons in air showers depends on the atmospheric density profile. This dependence is most prominent for high-energy ($\mathcal{O}$(TeV)) muons, mainly due to the decay length of charged pions and kaons becoming comparable to their interaction lengths. We have explored this atmospheric dependence of the muon energy spectrum in air showers and how it depends on hadronic interaction models and the primary mass composition using \mceq.

In particular, we have considered how the number of high-energy muons in vertical cosmic-ray air showers of a fixed energy and mass is expected to vary throughout the year, and how this relates to e.g. the production altitude of the muons and the relative contributions from decays of pions and kaons. While such seasonal variations have been considered before~\cite{DeRidderICRC2013, Gaisser:2021cqh}, as well as included as a systemic uncertainty in more conventional cosmic-ray analyses~\cite{IceCube:2019hmk}, a measurement of this effect itself as a way to investigate shower physics and the primary cosmic-ray flux had not yet been described.

To assess this effect experimentally, one needs the combination of an air shower detector which can determine the primary energy of cosmic rays, and an underground detector which can observe high-energy muons from the corresponding air showers. Determining the relative variation of the muon number throughout the year may constitute a method of probing the cosmic-ray mass composition and hadronic interactions with potentially low systematic uncertainties. Such a measurement can currently be performed at the IceCube Neutrino Observatory, and may benefit also from the future expansion to IceCube-Gen2, including its planned surface array~\cite{IceCube-Gen2:2020qha}.

\section*{Acknowledgements}

We thank Anatoli Fedynitch and David Seckel for useful discussions related to this work. We acknowledge funding from the National Science Foundation (NSF) grants \#2046386 and \#2209483.

\bibliographystyle{JHEP}
{\small
\bibliography{main}
}

\end{document}